\documentclass{cpbtex}
\usepackage{graphicx,bm,bbm,color,amsmath}
\usepackage{geometry}
\usepackage{CJKutf8}
\usepackage{diagbox}
\usepackage{hhline}
\usepackage{color}
\begin{document}

\begin{CJK*}{UTF8}{gbsn}

\title{Neural Network Analytic Continuation for Monte Carlo:\\ Improvement by Statistical Errors}

\author{Kai-Wei Sun(孙恺伟)$^{1}$ and Fa Wang (王垡) $^{1,2}$\footnote{Email: wangfa@pku.edu.cn}\\
$^{1}$International Center for Quantum Materials, School of Physics,\\ Peking University, Beijing 100871, China\\
$^{2}${Collaborative Innovation Center of Quantum Matter, Beijing 100871, China}}
\date{\today}
\maketitle

\end{CJK*}

\begin{abstract}
This study explores the use of neural network-based analytic continuation to extract spectra from Monte Carlo data. We apply this technique to both synthetic and Monte Carlo-generated data. The training sets for neural networks are carefully synthesized without ``data leakage". We found that the training set should match the input correlation functions in terms of statistical error properties, such as noise level, noise dependence on imaginary time, and imaginary time-displaced correlations. We have developed a systematic method to synthesize such training datasets. Our improved algorithm outperform the widely used maximum entropy method in highly noisy situations. As an example, our method successfully extracted the dynamic structure factor of the spin-$\frac{1}{2}$ Heisenberg chain from quantum Monte Carlo simulations.
\end{abstract}



\section{Introduction}
Numerical analytic continuation (AC) solves the following inversion problem,
\begin{equation}
G(\tau)=\int d\omega K(\tau,\omega) A(\omega).
\end{equation}
The goal of AC is to extract the real frequency spectrum $A(\omega)$ from the imaginary-time correlation function $G(\tau)$, which is typically obtained by Monte Carlo simulation. The spectrum $A(\omega)$ is required to be non-negative at any $\omega$-point and subjected to certain sum rule $\int d\omega A(\omega)=\text{const}.$ $K(\tau, \omega)$ is the inversion kernel, the form of which varies on specific problems being handled. This study involves two kinds of inversion kernels $K(\tau,\omega)$ including $K_F(\tau,\omega)=e^{-\tau \omega}/(1+e^{-\beta \omega})$ and $K_S(\tau,\omega)=e^{-\tau \omega}$. $K_F(\tau,\omega)$ usually appears while calculating single-particle excitation spectra from measured Green's functions\cite{white1989monte,silver1990maximum}. $K_S(\tau,\omega)$ is usually involved while extracting dynamic structure factors from spin-spin correlation functions in some spin models\cite{henelius2000monte}. 

To carry out actual calculation, $\tau$ and $\omega$ are often discretized as $\tau_i=\tau_1,\cdots, \tau_M$, $\omega_i=\omega_1,\cdots, \omega_N$. Then the target problem can be reformulated as $G(\tau_i)=\sum_{j=1}^N K(\tau_i, \omega_j) A(\omega_j) \Delta \omega.$ For the purpose of simplicity, $\Delta \omega$ will be absorbed to $A(\omega_j)$ by $A(\omega_j)\Delta \omega \rightarrow A(\omega_j)$ in further discussions. The sum rule is then discretized to be $\sum_{i=1}^N A(\omega_i)\Delta \omega=\text{const}.$ It seems like a simple problem of matrix inversion at first sight but turns out to be a notoriously challenging task due to the ill-conditioned nature of this inversion problem. In almost all cases, corresponding condition numbers go far beyond the tolerance of existing computers' machine precision. Several methods are proposed to solve this problem such as the Maximum Entropy method (Maxent)\cite{silver1990maximum} and Stochastic Analytic continuation (SAC)\cite{sandvik1998stochastic}. Both of them succeed in extracting empirically correct spectra. However, these methods usually demand highly accurate simulated correlation functions $G_\text{sim}(\tau)$. 


As an emerging technique for machine learning, neural networks (NNs)\cite{gurney2018introduction} have experienced great success in a variety of physics-related domains. From the perspective of machine learning, analytic continuation can be viewed as a vector-to-vector prediction task, where $G(\tau_i)$ is mapped to $A(\omega_j)$. To construct a neural network capable of performing analytic continuation, both the network topology and training set should be built appropriately. The common framework on this task usually contains several steps: (1) Build a neural network. (2) Synthesize spectra $A_\text{train}$ for training purpose. (3) Calculate $G_\text{train}$ by the forward mapping $A \rightarrow G$. Noting that the forward mapping is well-conditioned, thus $G_\text{train}$ can be exactly determined. (4) Train the network using the dataset pair $(G_\text{train}, A_\text{train})$ so that spectra predicted from $G_\text{train}$ closely match $A_\text{train}$. (5) When developing and testing NNs, synthesize testing set ($G_\text{test}$, $A_\text{test}$) and evaluate the performance of trained NN on it. When using NNAC in actual tasks, apply trained network to predict spectra $A_\text{pred}$ from simulated correlation functions $G_\text{sim}$ generated by Monte Carlo simulations. To mimic real-world simulated data, noises are usually added to correlation functions obtained from synthetic spectra such as $G_\text{train}$ and $G_\text{test}$.

In a relatively early study, Hongkee Yoon\cite{yoon2018analytic} and co-authors designed a network mainly based on fully-connected-layers (FCLs)\cite{gurney2018introduction}. In their research, both training and testing sets are obtained from synthetic Gaussian-type multi-peak spectra. Noises of Gaussian distribution are added to $G_\text{train}$ and $G_\text{test}$. The trained NN performs well in the testing set as the predicted spectra are very close to synthetic testing spectra. Several different network structures\cite{arsenault2017projected,xie2021analytic,huang2022learned,zhang2022training} trained on similar Gaussian-type datasets are also proposed. In addition to synthetic datasets, neural networks based analytic continuation (NNAC) are also examined on some exactly solvable models such as one-dimensional transverse-field Ising model\cite{yao2022noise} and harmonic oscillator linearly coupled to an ideal environment\cite{fournier2020artificial}. In these two studies, artifitial training sets ($G_\text{train}$,$A_\text{train}$) are generated from exactly solved correlation functions and corresponding spectra. Different spectra in the training set correspond to different parameter values in the Hamiltonian being studied. Target spectra $A_\text{pred}$ are predicted from simulated correlation function $G_\text{sim}$ using Monte Carlo techniques. Ref\cite{yao2022noise} points out that the neural network's prediction performance can be improved by adding uniform noises to the exactly solved Green's functions at each imaginary time in the training set.

Theoretically we have no knowledge about precise forms of spectra to be predicted before target spectra are actually predicted. That's because the knowledge of Gaussian-type spectra are not expected to be known before prediction. This is actually an intriguing topic dubbed ``data leakage"\cite{kaufman2012leakage} in the field of machine learning. Data leakage occurs when information is used in the training process but not expected to be available at prediction time. All aforementioned articles about NNAC have the issue of data leakage at some levels. In practice, we usually apply numerical analytical continuation to models that are not exactly solvable, where it is not possible to construct training sets by exactly solved spectra.

To design the training set, hints from experiments or traditional AC approaches such as Maxent should also be explored. It should be mentioned that NNAC is useful even when spectra are already obtained from Maxent: NNAC performs better at least in highly-noisy cases as described in Ref\cite{fournier2020artificial}. This topic will also be elaborated upon in this paper. In general, domain knowledge\cite{jordan2015machine,domingos2012few}, especially possible spectrum peak shapes, should be incorporated when designing the training set as much as feasible but without data leakage. We then expect the trained NN to generalize\cite{giles1987learning,novak2018sensitivity} well enough to handle unobserved correlation functions like $G_\text{test}$ and $G_\text{sim}$.

Intuitively, people expect better prediction of spectra by incorporating more information. Monte Carlo simulations can provide more information beyond the measured correlation functions, such as the statistical errors of $G(\tau)$. Specifically, they can provide information regarding two aspects of statistical errors: the measured errors $R(\tau_i)$ of $G(\tau_i)$ at each $\tau_i$, and the covariance of correlation functions at different imaginary times.

This work avoids data leakage while synthesizing the training sets and incorporates information of statistical errors to improve the performance of NNAC. With these means, NNAC has the potential to be a usable algorithm in practical applications and a significant component in the Monte Carlo-Analytic Continuation toolchain. In section 2, NNAC of kernel $K_F(\tau,\omega)$ is examined on synthetic data, where datasets synthsized from spectra of different types of shapes are addressed. In section 3, NNAC of kernel $K_S(\tau,\omega)$ is applied to one-dimensional Heisenberg chain as a real-world example of an AC problem. Conclusions are presented in the final section.

\section{NNAC on Synthetic Datasets}
In this section, we design and test NNs on synthetic datasets. Principles for generating training sets will be developed. We first discuss three types of datasets, the training framework, as well as the actual training process. Noise level matching between the training and the testing set is then explored. Resulting spectra are then compared with those from Maxent. The impact of measured noise shapes and time-displaced correlation is then investigated. 
\subsection{Preparation of Dataset}
Multi-peak spectra $A(\omega)$ are produced by summing over single peaks $F(\omega)$.
\begin{equation}
    A(\omega)=\frac{1}{Z} \sum_i F_i(\omega).
\end{equation}
In the formula above, $Z$ is a scaling constant ensuring that $A(\omega)$ obeys the sum rule. In this section, we assume $\int d \omega A(\omega)=1$ for convenience. This paper involves three distinct peak types: asymmetric exponential power(ASEP), skew Gaussian(Skew), and Lorentz. The ASEP single-peak curve reads:
\begin{equation}
F^\text{ASEP}(\omega)=\left \{
\begin{aligned}
&h  \exp \big [-(\frac{m-\omega}{a_1})^{b_1} \big ], \omega<m ; \\
&h  \exp \big [-(\frac{\omega-m}{a_2})^{b_2} \big ],\omega\geq m .
\end{aligned}
\right.
\end{equation}
In the above formula, $h$, $m$, $a_1$, $a_2$, $b_1$, $b_2$ are all control parameters. In this study, we set $m\in [-5,5]$, $a_1,a_2 \in [0.3,3]$, $b_1,b_2 \in [1,3]$, $h \in [0.2,1]$. The Skew peak takes the form

\begin{equation}
F^\text{Skew}(\omega)=\left \{
\begin{aligned}
&0  , z\leq 0; \\
&\frac{h}{az}\exp(-\frac{y^2}{2}),z>0.
\end{aligned}
\right.
\end{equation}
$z(\omega)= 1-k\frac{\omega-m}{a}$ and $y=\frac{1}{k}\ln(z)$. Control parameters are $m\in [-2,2]$, $a \in [0.5,1]$, $k \in [-1,1]$, and $h \in [0.2,1]$. The Lorentz curve takes the relatively simple form
\begin{equation}
    F^\text{Lorentz}(\omega)=h\frac{1}{(\omega^2-a^2)^2-g^2\omega^2},
\end{equation}
where $g \in [1,2]$, $a \in [2,4]$ and $h \in [0.2,1]$. In this study, we investigate spectra containing one to four peaks. At least $10^5$ samples are generated for each peak number by randomly selecting control parameters. In other words, one single dataset includes at least $4\times 10^5$ samples. Training and testing sets of the same peak type are independently created. 

ASEP-type dataset has the most control parameters among these three types and thus contains a greater diversity of spectra while not explicitly contain spectra of Skew-type or Lorentz-type dataset. We expect the neural network to learn from ASEP-type dataset and generalize effectively to achieve good performance on the other two datasets. It should be noted that, unlike in some previous studies, we will not examine Gaussian-type spectra here, as they are explicitly included in ASEP-type dataset when $b_1=b_2=2$ and $a_1=a_2$. This explicit inclusion case does not frequently occur in real-world AC tasks and the performance of NNAC will be overestimated in the case of Gaussian-type testing sets.

The imaginary time $\tau \in [0,16]$ is discretized uniformly into 512 pieces and the frequency domain $\omega \in [-15,15]$ is discretized into 1024 pieces. $\beta$ is fixed to be $16$ in the Fermion kernel $e^{-\tau \omega}/(1+e^{-\beta \omega})$.

\subsection{Training Framework}
\begin{figure}[htbp]
\centering
\includegraphics[width=0.7 \linewidth]{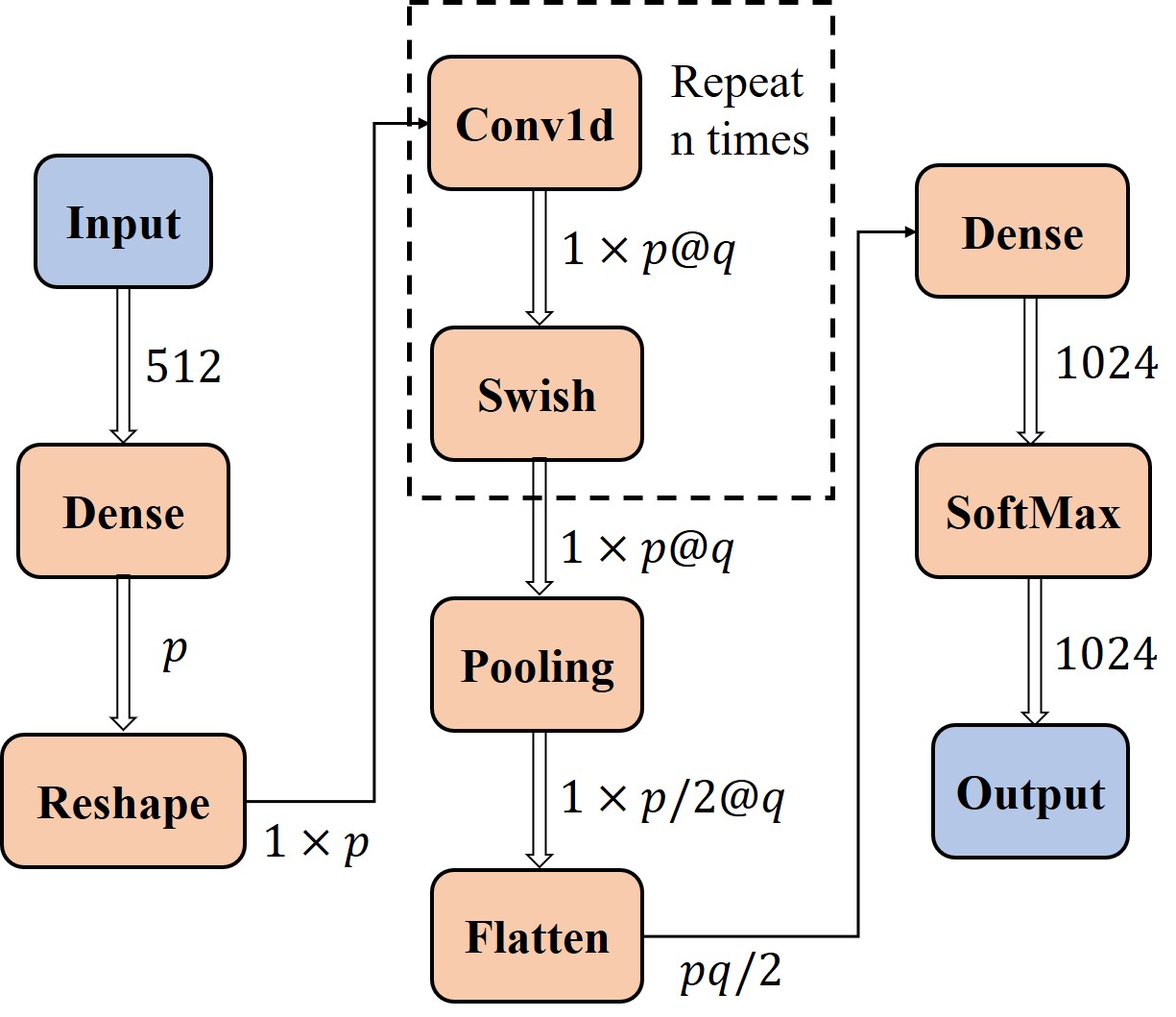}
\caption{The convolution-based structure of the neural network used in this work. Hyper-parameters are chosen to be $n=8$, $p=64$ and $q=64$ in actual training process.}
\label{net_cnn}
\end{figure}

Convolution neural networks (CNNs)\cite{albawi2017understanding} will be employed in this work. FCL-based neural networks are also evaluated in the early stage of this study, which proves inferior to CNNs. Involvement of residual modules\cite{he2016deep} or deep layer aggregation\cite{yu2018deep} also does not prove to make significant improvements. In the case of deep layer aggregation, both iterative deep aggregation and hierarchical deep aggregation are attempted. Based on the aforementioned factors, we employ the neural network shown in Figure \ref{net_cnn}. At first the 512-length $G(\tau_i)$ is transferred to a $p$-length vector via a FCL (labeled ``Dense") and then reshaped to be a $1 \times p$ matrix. This matrix can be regarded as a specific image that can be naturally processed by convolution layers. Next, this image is passed to a $q$-channel one dimensional convolution layer ``Conv1d", followed by the activation layer ``Swish". Within the ``Conv1d" layer, convolution kernels of size $1\times3$ are used. Within the activation layer, the activation function named ``Swish"\cite{ramachandran2017searching} is used. This activation function is both non-monotonic and smooth and may improve the overall performance of the neural network compared to the commonly used ReLU\cite{agarap2018deep} activation function according to Ref\cite{ramachandran2017searching}. This ``convolution $\rightarrow$ activation" process will be carried out $n$ times. The $q$-channel image is then compressed by an average-pooling layer\cite{iosifidis2022deep} and flattened to be a $pq/2$-long vector. The flattened vector will be mapped to a 1024-long vector by another ``Dense" layer. Ultimately, the ``SoftMax" layer will present predictions of the spectra where the sum rule $\sum_j A(\omega_j)=1$ is naturally satisfied after this softmax operation. Tricks to reduce overfitting such as dropout\cite{srivastava2014dropout} are not adopted here. Instead, we recommend enlarging the training set when signs of overfitting emerge since it is rather cheap to acquire data from synthetic spectra.

Hyper-parameters are chosen to be $n=8$, $p=64$, and $q=64$. To select appropriate hyper-parameters, we build an additional ASEP-type validation set, on which to evaluate NN trained by ASEP-type training set. When selecting hyper-parameters, the trade-off between performance and training time is taken into consideration. 

We use Kullback-Leibler Divergence(KLD)\cite{joyce2011kullback} as the loss function, which takes the form
\begin{equation}
D_\text{KL}(A_\text{true} || A_\text{pred})=-\sum_{j} A_\text{true}(\omega_j) \ln\frac{A_\text{true}(\omega_j)}{A_\text{pred}(\omega_j)}.
\end{equation}
KLD measures the difference (more precisely, relative entropy) between the true distribution $A_\text{true}$ and the predicted distribution $A_\text{pred}$, which makes it a natural choice in this task. Other commonly-used loss functions include mean absolute error (MAE) and mean squared error (MSE) as shown below. KLD also has the property of positivity as MAE and MSE.
\begin{align}
    \text{MAE}(A_\text{true},A_\text{pred})&=\frac{1}{N}\sum_{j=1}^N \big |A_\text{true}(\omega_j)-A_\text{pred}(\omega_j) \big|\\
    \text{MSE}(A_\text{true},A_\text{pred})&=\frac{1}{N}\sum_{j=1}^N \big [A_\text{true}(\omega_j)-A_\text{pred}(\omega_j) \big ]^2
\end{align}

Empirically, spectra from NNs with MSE loss are often smoother than those from NNs with MAE loss since MSE punish large spectrum difference more severely. In this study, we didn't observe discernible difference in the performance between MSE-loss and KLD-loss NNs.

NNs are programmed using Keras toolkits\cite{chollet2015keras} with Tensorflow\cite{tensorflow2015-whitepaper} backends. The Adam\cite{kingma2014adam} optimizer is used for gradient descent. The early-stopping trick is utilized during training. The training process terminates when KLD measured on the validation set does not drop for 20 epochs, where the validation set is generated in the same manner as the training set. Trained weights are then restored to the epoch with the lowest KLD. Each training task will be repeated at least 5 times with different random seeds. KLDs shown in this paper are averaged among NNs trained with different seeds.

\begin{figure}[htb]
\centering
\includegraphics[width=0.9 \linewidth]{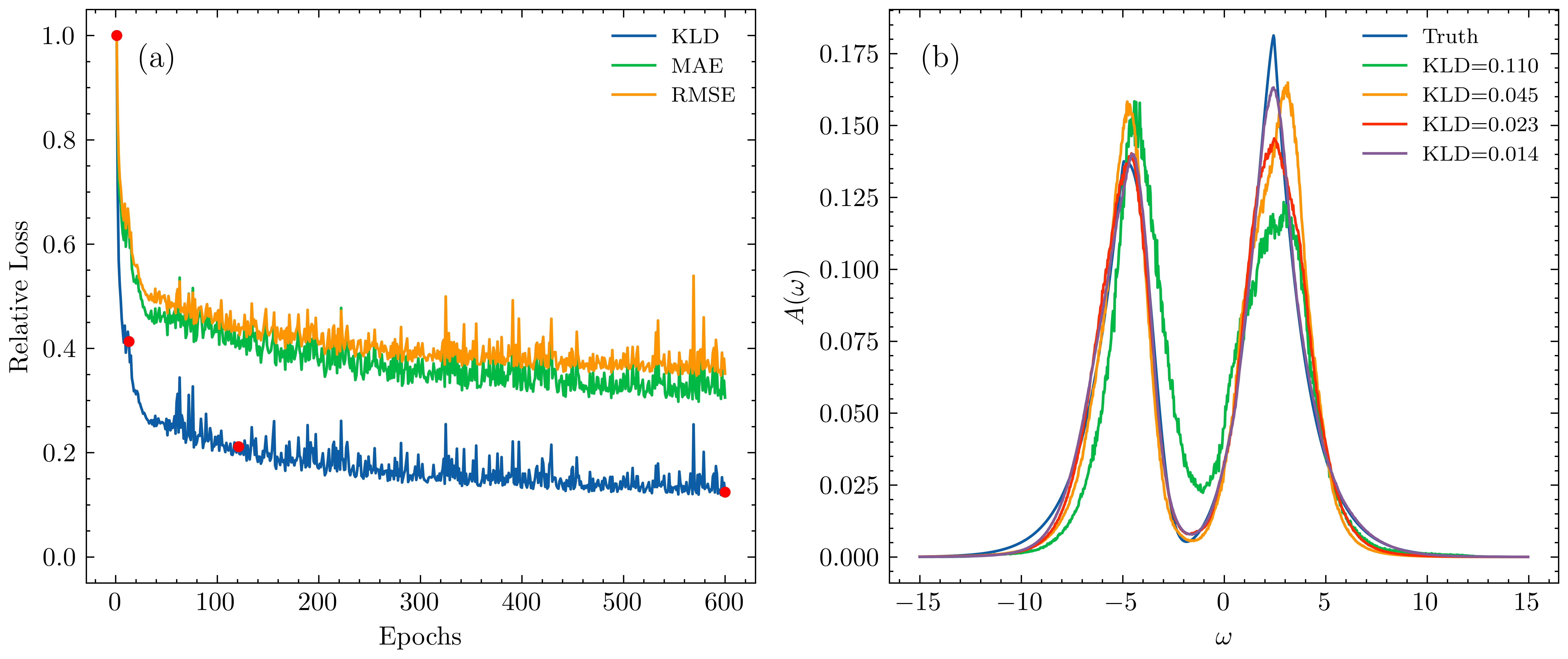}
\caption{Tracking the training process. (a) Relative losses, including KLD, MAE, and RMSE, \textit{w.r.t.} number of trained epochs. This so-called relative loss is defined as ``loss after this epoch"/``loss after the first epoch". (b) A typical example of the convergence process of one predicted spectrum to the true spectrum as KLD decreases. Selected checkpoints are labeled by red dots in (a).}
\label{process}
\end{figure}

The training process is depicted in Figure \ref{process}, where both the training set and the testing set are of ASEP-type. Errors at noise level $10^{-3}$ are introduced to $G_\text{train}$ and $G_\text{test}$ (the concept of noise level will be discussed later). Relative values of three statistics measured on the testing set are tracked throughout the training process in Figure \ref{process} (a). We track $\text{RMSE}=\sqrt{\text{MSE}}$ instead of $\text{MSE}$ itself because RMSE shares the same dimension as MAE and KLD. Relative loss in this figure is defined as ``loss after this epoch"/``loss after the first epoch". In Figure \ref{process} (b) we show an example from the testing set of how one predicted spectrum becomes closer to the true spectrum at different KLD levels. Selected checkpoints are indicated by red dots in Figure \ref{process} (a). While visualizing the training process, we only use 1000 samples for each epoch because statistics converge too quickly for visualization if the entire training set containing $4 \times 10^5$ samples is used. The complete training set will be used in actual AC tasks hereafter.

In this study, model training on an RTX3060 graphics card takes approximately 20 minutes on average. This is acceptable in the majority of circumstances, especially in contrast to the amount of time saved in the Monte Carlo simulation if highly accurate correlation functions are not incorporated.

\subsection{Noise Level Matching}
Correlation functions measured from Monte Carlo simulation inevitably contain statistical errors. To mimic simulated errors, Gaussian-type noises are added to $G(\tau_i)$ by $G(\tau_i) \rightarrow G(\tau_i)+R(\tau_i)$, where $R(\tau_i) \sim N(0,\sigma^2)$. Four different noise levels are investigated in this work, $\sigma=10^{-4}, 10^{-3}, 3\times 10^{-3}, 10^{-2}$. $\sigma$ in this formula can also be interpreted as the absolute average of noises. At this stage, we assume $G(\tau_i)$ to be independently measured for each $i$. In real-world NNAC-based tasks, noises of $G_\text{sim}$ are measured from Monte Carlo simulation, and noises of the training set should be carefully arranged accordingly. Besides, the noise level of the testing set should be the same as the simulated data to mimic real-world tasks.

\begin{figure}[htbp]
    \centering
    \includegraphics[width=0.9 \linewidth]{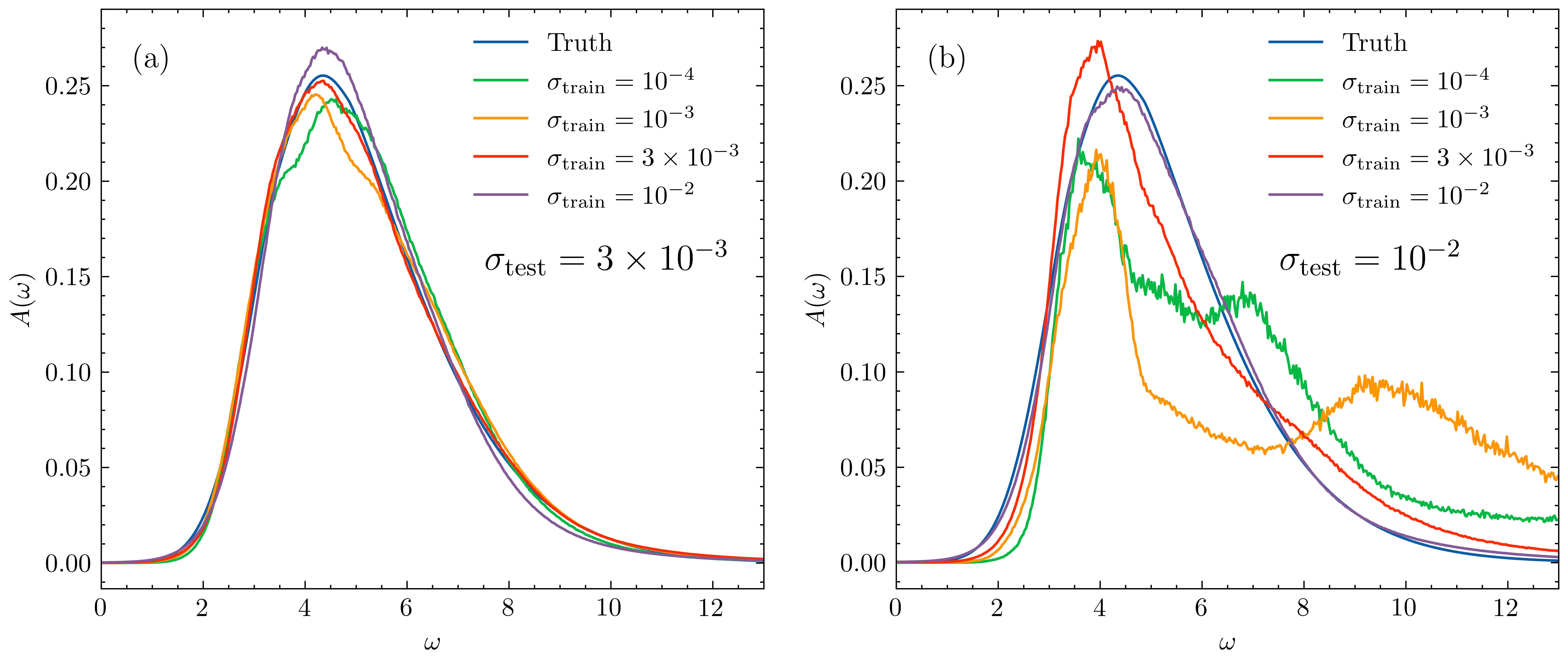}
    \caption{Illustration of noise level matching. Ground truths in both sub-figures are the same curve. (a) Prediction of spectra from testing set with $\sigma=3\times 10^{-3}$ by NNs trained with different $\sigma_\text{train}$. The best spectrum is obtained when $\sigma_\text{train}=\sigma_\text{test}=3\times 10^{-3}$. (b) Prediction of spectra from testing set with $\sigma=10^{-2}$ by NNs trained with different $\sigma_\text{train}$. The best spectrum is obtained when $\sigma_\text{train}=\sigma_\text{test}=10^{-2}$. The predicted spectrum contains unstable peaks at wrong locations when $\sigma_\text{train}=10^{-4}$ or $3\times 10^{-3}$.}
    \label{mutual_noise}
    \end{figure}

To design the training set, a natural question arises as how we should set noise level $\sigma_\text{train}$ of the training set when the noise level $\sigma_\text{test}$ of the testing set is known? We train NNs by training sets of different $\sigma_\text{train}$ and apply these NNs on testing sets of different $\sigma_\text{test}$. Corresponding results are shown in Table \ref{tab_mutual_noise} and Figure \ref{mutual_noise}. Table \ref{tab_mutual_noise} contains KLDs of spectra predicted from testing sets with different noise levels $\sigma_\text{test}$ by NNs trained by training sets with different $\sigma_\text{train}$. The smallest KLD in each line (marked red) is obtained when noise levels of the training set and the testing set match ($\sigma_\text{train}=\sigma_\text{test}$). Performance degrades but remains acceptable when $\sigma_\text{train}$ increases and $\sigma_\text{train}>\sigma_\text{test}$ while the opposite is not true when $\sigma_\text{train}<\sigma_\text{test}$. For instance, KLD is relatively small when $(\sigma_\text{train},\sigma_\text{test})=(10^{-2},10^{-4})$ but is large and unsatisfactory when $(\sigma_\text{train},\sigma_\text{test})=(10^{-4},10^{-2})$. That's because information of ASEP($\sigma=10^{-4}$) is somehow ``contained'' in ASEP($\sigma=10^{-2}$): for each curve in ASEP($\sigma=10^{-4}$) we may find similar samples with similar noises in ASEP($\sigma=10^{-2}$) if datasets are large enough given noises are randomly selected, whereas the converse is not true. We train NNs with different noise levels and use them to predict one sample of $G(\tau_i)$ from the testing set with $\sigma_\text{test}=3\times 10^{-3}$ and $\sigma_\text{test}=10^{-2}$, which are presented in Figure \ref{mutual_noise} (a) and (b), respectively. Resulted spectra become closer to the ground truth when $\sigma_\text{train}$ is closer to $\sigma_\text{test}$. In Figure \ref{mutual_noise} (b), incorrect and unstable peaks are predicted by the NNs trained with $\sigma_\text{train}=10^{-4}$ or $10^{-3}$, whose KLDs are large correspondingly as seen in Table \ref{tab_mutual_noise}.

Note that in this part, data leakage is not intentionally avoided: the training set and the testing set are both of ASEP type. With the same $\sigma_\text{test}$, KLD differences caused by different $\sigma_\text{train}$ may be relatively small and taking datasets with different line shapes may introduce unnecessary complexity, resulting in unsolid or even incorrect conclusions. From another perspective, we expect NNs to use the knowledge learned from the training set to predict correct spectra in actual tasks. The performance will be usually slightly weakened if line shapes of the testing set and training set are different. Therefore, we expect the NNs 
of proper $\sigma_\text{train}$ to at least achieve good results on the testing set with the same line shape. The KLD results here do not represent actual performances of the NNs in practical tasks.

\begin{table}[htbp]
    \centering
    \begin{tabular}{||l|l|l|l|l||} 
    \hhline{|t:=====:t|}
    \diagbox{$\sigma_\text{test}$}{$\sigma_\text{train}$} & $10^{-4}$ & $10^{-3}$ & $3\times10^{-3}$ & $10^{-2}$\\ 
    \hhline{|:=====:|}
    $10^{-4}$    & \textcolor{red}{0.0137(3)} & 0.0151(4) & 0.0181(2)  & 0.0280(1)  \\ 
    \hhline{|:=====:|}
    $10^{-3}$    & 0.0172(1) & \textcolor{red}{0.0164(4)} & 0.0185(2)  & 0.0280(1)  \\ 
    \hhline{|:=====:|}
    $3\times10^{3}$   & 0.045(2) & 0.0268(3) & \textcolor{red}{0.0217(1)} & 0.02854(9)  \\ 
    \hhline{|:=====:|}
    $10^{-2}$  & 0.31(2) & 0.148(6) & 0.060(1)  & \textcolor{red}{0.0350(1)}  \\
    \hhline{|b:=====:b|}
    \end{tabular}

\caption{KLDs of spectra predicted from testing sets with different$\sigma_\text{test}$ by NNs trained by training sets with different $\sigma_\text{train}$. In each line, the smallest KLD (marked red) is obtained when $\sigma_\text{train}=\sigma_\text{test}$. To determine the errors of the KLDs in the table, we train NNs with at least 10 distinct random seeds and calculate statistical uncertainty of KLDs of spectra predicted by these NNs.}
\label{tab_mutual_noise}
\end{table}

\subsection{Comparison with Maxent}
\begin{figure}[htbp]
\centering
\includegraphics[width=0.9 \linewidth]{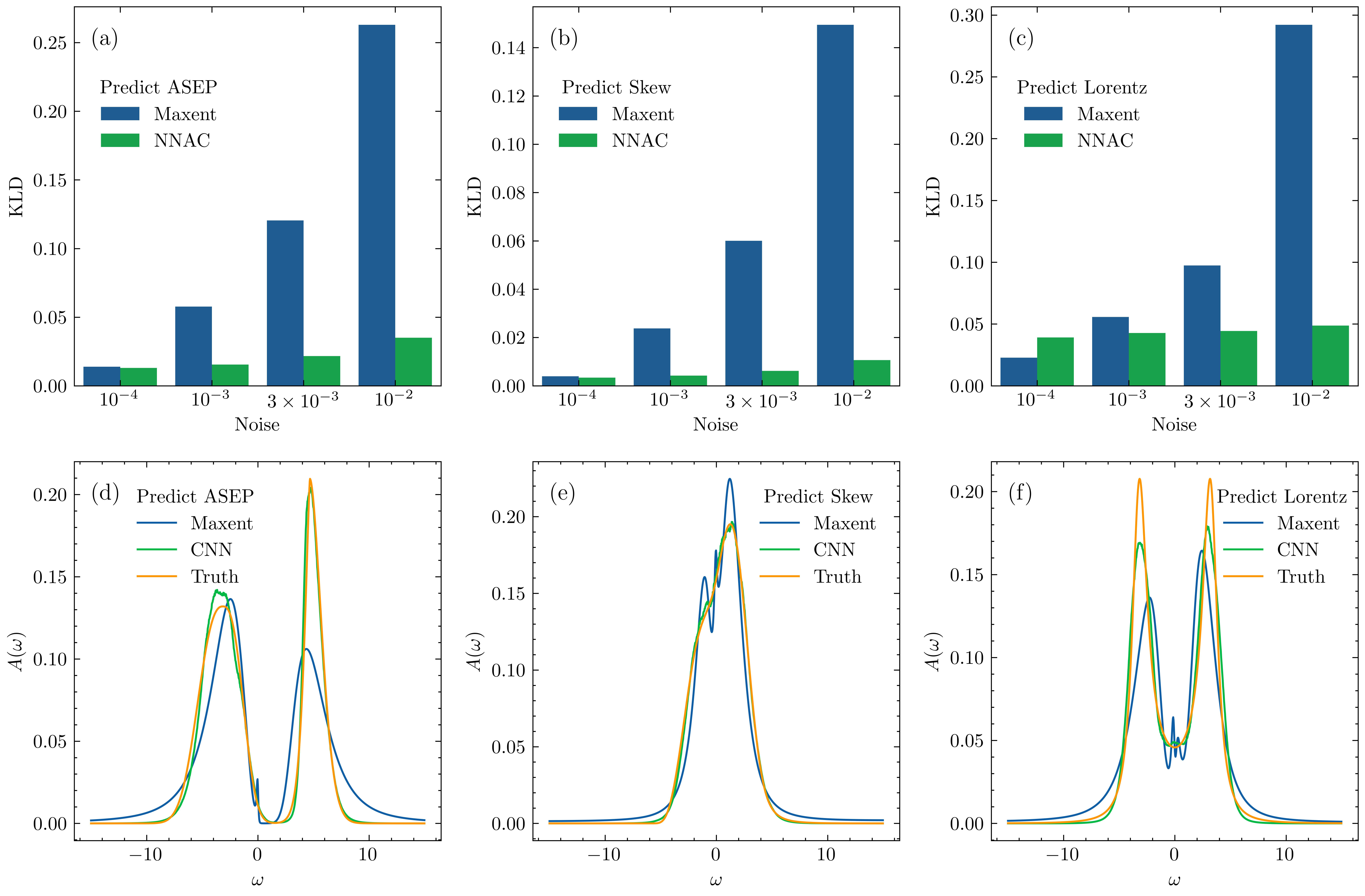}
\caption{Comparison with Maxnet. NNs are trained by ASEP dataset and applied on three different testing sets: ASEP, Skew, and Lorentz. (a) to (c): KLD predicted results of ASEP, Skew, Lorentz dataset respectively at different noise levels. (d) to (f): typical predicted spectra at the noise level $3 \times 10^{-3}$ by Maxent and NNAC. The ground truth is also shown as comparison. The performance of NNAC is comparable with Maxent when the dataset contains low-level noise but outperforms Maxent at high-level noise even if NNs are not trained by the dataset of the same type as the testing set.}
\label{compare_maxent}
\end{figure}

With the knowledge of noise level matching, $G_\text{train}$ will be designed to have the same noise level as $G_\text{test}$ in this work hereafter and we are now ready to compare NNAC with traditional AC methods like Maxent. We train NNs by ASEP training sets and use them to predict ASEP-type, Skew-type and Lorentz-type spectra, respectively. Corresponding outcomes are depicted in Figure \ref{compare_maxent}. Figure \ref{compare_maxent} (a),(b) and (c) show KLDs of spectra predicted by these two methods on ASEP, Skew, and Lorentz dataset respectively. Error bars of KLDs are omitted in this and subsequent figures to make graphs more comprehensible as they are relatively small. Typical predicted results at noise level $3 \times 10^{-3}$ are shown in Figure \ref{compare_maxent} (d),(e) and (f) of three peak types. Performance of NNAC is comparable with Maxent at the lowest noise level $10^{-4}$ but outperforms Maxent significantly at relatively high noise levels. The improvement of prediction effect is also obvious when the training set and testing set are not of the same spectrum type. In spectrum examples depicted in Figure \ref{compare_maxent} (d),(e) and (f), peak locations are precisely predicted by NNAC but Maxent didn't provide accurate peak locations at this noise level. In some frequencies, Maxent may even give incorrect signals of peaks. Peak heights predicted by NNAC are also more accurate and closer to ground truths than Maxent's.

Spectra from Maxent in this section about kernel $K_F(\tau,\omega)$ are calculated mainly based on the software ``TRIQS/maxent''\cite{PhysRevB.96.155128} so that results can be easily checked. Various $\alpha$-choosing algorithms are evaluated, where $\alpha$ is the penalty coefficient of the entropy term in the Maxent objective function\cite{silver1990maximum}. Among these algorithms discussed in Ref\cite{PhysRevB.96.155128}, ``$\chi_2$-curvature'' , which is analogous to $\Omega$-Maxent\cite{PhysRevE.94.023303}, and ``Bryan'' algorithms greatly outperform others in terms of KLD in tasks of interest. Between these two, ``$\chi_2$-curvature'' is marginally superior to the Bryan algorithm. In this way, we use ``$\chi_2$-curvature'' in this work to ensure a level playing field for Maxent.

\subsection{Influence of Noise Dependency on Imaginary Time}
\begin{figure}[htbp]
\centering
\includegraphics[width=0.7 \linewidth]{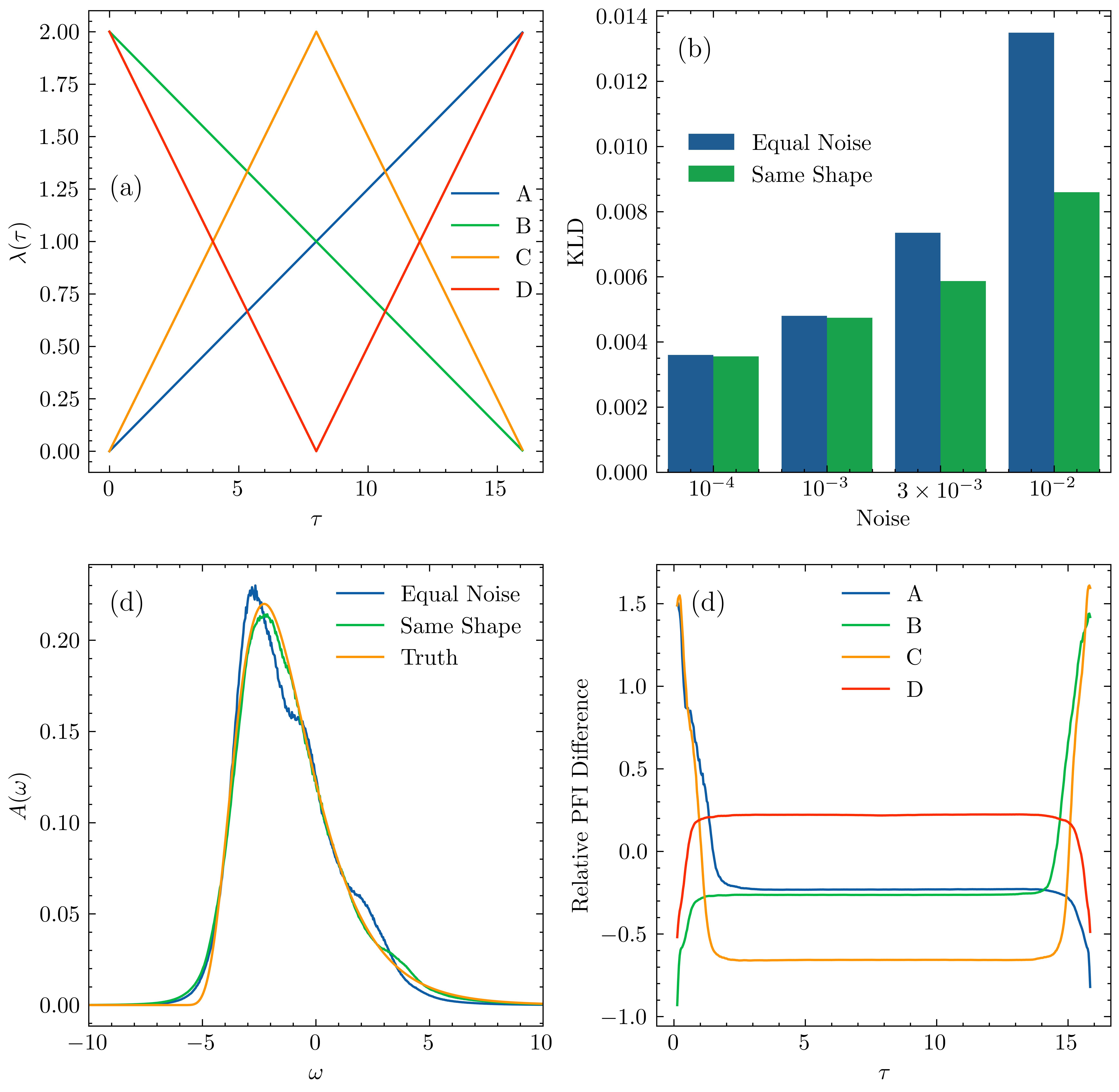}
\caption{Influence of linear noise shapes. (a) Four types of shape multiplier $\lambda(\tau)$. (b) Noises of the testing set are of shape A. Two neural networks are trained by training set of equal noise ($\lambda(\tau)=1$) and noise shape A, respectively. KLDs of trained neural networks are compared on the shape-A testing set at different noise levels. (c) Typical spectra predicted from $G(\tau)$ of shape-A noises ($\sigma=3\times10^{-3}$) by neural networks trained by equal-noise and shape-A training sets, respectively. (d) The relative difference in PFI between the neural network trained on a training set with linearly-shaped noise and the neural network trained on a training set with uniformly-shaped ($\lambda(\tau)=1$) noise at various imaginary times.}
\label{linear_noise}
\end{figure}

In the preceding discussion, noise $R(\tau_i)$ at each $\tau_i$ is assumed to be sampled from the same Gaussian distribution and has the same variance, which is rarely the case in Monte Carlo simulation. We introduce the noise-shape-multiplier $\lambda(\tau)$ to investigate influence of noise dependency on imaginary Time and assume $R(\tau_i) \sim N(0,\sigma(\tau_i)^2)$, $\sigma(\tau_i)=\lambda(\tau_i) \sigma$. We refer to this dependency as "noise shape" hereafter. These multipliers satisfy $\frac{1}{\beta}\int_0^\beta \lambda(\tau) d\tau=1$ to ensure that datasets with the same $\sigma$ but different noise shapes are at approximately the same noise level. $\lambda(\tau)$ of four distinct linear shapes labeled A, B, C, and D are displayed in Figure \ref{linear_noise} (a). 

To demonstrate the impact of noise shape and how to appropriately arrange noises in the training set, we train NNs by ASEP-type training sets with equal noise ($\lambda(\tau) =1$) and noise shape A, respectively. These trained NNs are implemented on Skew-type testing sets with noise shape A. Corresponding measured KLDs are presented in Figure \ref{linear_noise} (b). Spectra examples at noise level $3\times 10^{-3}$ are shown in in Figure \ref{linear_noise} (c).

Origins of different performances by different noise shapes can be, to some extent, explained by permutation feature importance (PFI)\cite{altmann2010permutation}, despite the fact that neural networks are typically seen as black boxes. To calculate PFI, we rearrange $G(\tau_i)$ randomly over samples on one certain time piece $\tau_i$ in the testing set and PFI at this time piece is defined by how much the resulted KLD increases. PFI difference between NNs trained by datasets of linear noise shapes and equal-noise dataset are defined by $[\text{PFI}^\text{T}(\tau_i)-\text{PFI}^\text{E}(\tau_i)]/[\text{PFI}^\text{T}(\tau_i)+\text{PFI}^\text{E}(\tau_i)]$. $\text{PFI}^\text{E}(\tau_i)$ denotes PFI from NNs trained by equal-noise dataset and $\text{PFI}^\text{T}(\tau_i)$ denotes PFI from NNs trained by dataset of some other noise shape, where $\text{T} \in [\text{A},\text{B},\text{C},\text{D}]$. Resulted relative PFI differences are shown in Figure \ref{linear_noise} (d). Moving average of adjacent five points are carried out to make curves smoother and clearer. Relative PFI curves and $\lambda(\tau)$ curves increase or decrease in the opposite direction, which means NNs assign large feature importance on imaginary time pieces where $G(\tau_i)$ are less noisy.

It should be emphasized that measured correlation functions do not often have linear-type noise shapes. Instead, they are frequently of exponential-like shapes. However, things can become more subtle in the case of exponential noise shape, when it becomes more difficult to disentangle the effects of different noise levels and noise shapes. In light of these concerns, we only examine linear-type noise shapes here, and it is believed that physical images are similar in other scenarios.

\subsection{Influence of Time-Displaced Correlation}
So far we've assumed that $G(\tau_i)$ at different $\tau_i$ are measured independently, which is not always true in practical Monte Carlo simulation. At this time, covariance instead of independent errors of $G(\tau_i)$ should be considered. Covariance can be decomposed as $\Sigma=U^TCU$. $U=[U(\tau_1),\cdots,U(\tau_N)]^T$, where $U(\tau_i)$ is the independently measured statistical error of $G(\tau_i)$. $C$ is the correlation matrix. $C_{ij}$ is defined as Pearson correlation of measured $G(\tau_i)$ and $G(\tau_j)$. In practical AC tasks, $\Sigma$ of $G_\text{sim}$ should be measured before designing the training set. If we require the training set to share the same covariance as the testing set, noises of the training set should be generated from corresponding joint Gaussian distribution, that is, $R \sim N(0,\Sigma)$.

To illustrate influences of time-displaced correlation, we create the toy correlation matrix for the testing set by 
\begin{equation}
    C_{ij}=\frac{1}{1+|i-j|^{1/\gamma}}.
\end{equation}
In this work, we will investigate correlation matrices with condition numbers being $10^3$, $10^6$, $10^9$, and $10^{12}$ respectively by adjusting $\gamma$. $U(\tau)$ are generated at four noise levels $\sigma \in [10^{-4}, 10^{-3}, 3 \times 10^{-3}, 10^{-2}]$.

\begin{figure}[htbp]
\centering
\includegraphics[width=0.9 \linewidth]{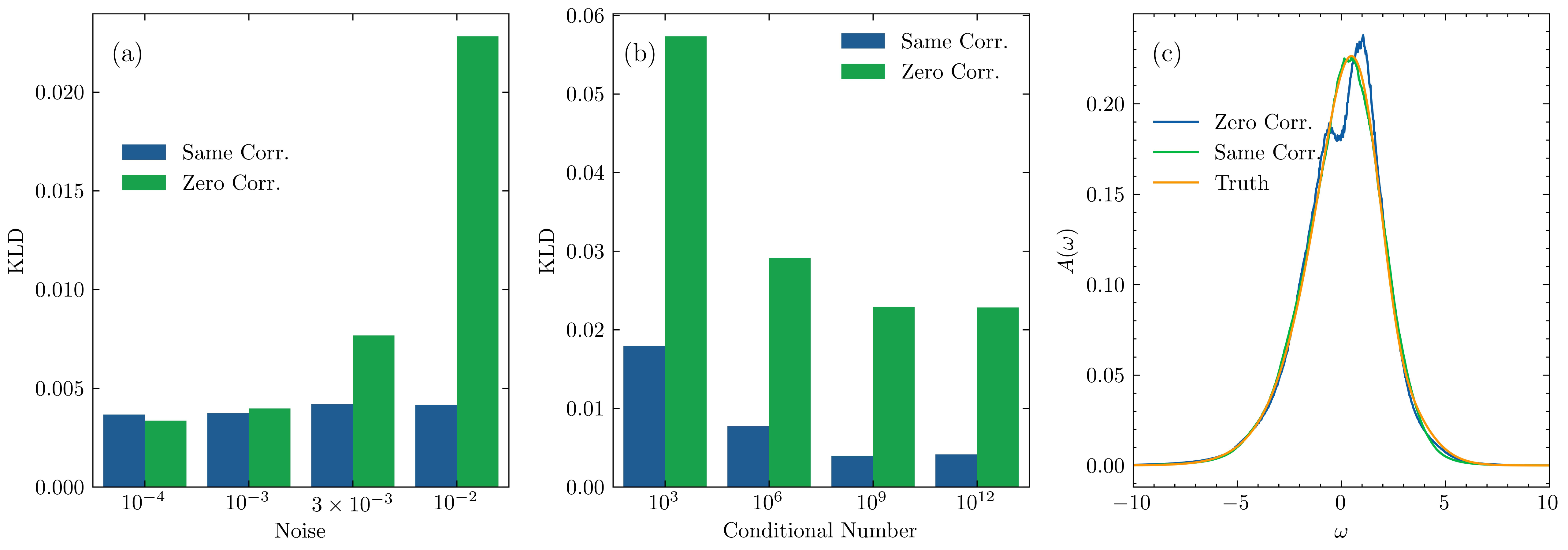}
\caption{Illustration of influences of time-displaced correlation of $G(\tau_i)$. Noise levels of the training set and the testing set are matched. (a) The condition number of the correlation matrix is fixed to be $10^{12}$. NNs trained by datasets without correlation may give wrong predictions if $G(\tau_i)$ in the testing set is correlated, especially when the noise level is high. (b) Noise levels are fixed to be $10^{-2}$. KLDs are shown \textit{w.r.t.} different condition numbers. (c) Spectra examples with condition number $10^{12}$ and noise level $10^{-2}$.}
\label{cov}
\end{figure}

NNs are trained by ASEP-type datasets and are to be applied to Skew-type testing sets with various noise levels and condition numbers. Training sets are designed in two manners: they may have zero correlation or the same correlation as the testing set. In Figure \ref{cov} (a), condition number of the testing set is fixed to be $10^{12}$. NNs are trained by dataset with or without time-displaced correlation on each noise level. As being illustrated, influence of $\tau$-correlation is not significant at low noise levels but correlation mismatching may lead to incorrect prediction at high noise levels. In Figure \ref{cov} (b), the noise level of the testing set (and the training set, as well) is fixed to be $10^{-2}$, where KLDs are lower when condition number is smaller. The reason may be that $R(\tau_i)$ are dominated by only a few singular values of $\Sigma$, whose pattern of noises is relatively easy to be learned by NNs. Spectrum examples are shown in Figure \ref{cov} (c) with noise level $10^{-2}$ and condition number $10^{12}$, which contain predicted spectra by NNs trained with zero or the same correlation as the testing set, as well as the ground truth. Clearly the predicted spectra contain wrong peaks at wrong locations when time-displaced correlation is not matched.

\section{NNAC on Heisenberg Chain}
In this section, NNAC is carried out to extract dynamic structure factors of the spin-$\frac{1}{2}$ anti-ferromagnetic Heisenberg chain of length $L$, which reads
\begin{equation}
    H=J\sum_{i=1}^L \vec{S}_i \cdot \vec{S}_{i+1}.
\end{equation}
$\vec{S}_i$ represents a spin located on site $i$. Periodic boundary condition is assumed, \textit{i.e.}, $\vec{S}_{L+1}=\vec{S}_1$. Imaginary-time-displaced spin-spin correlation of $z$-component is measured by stochastic series expansion\cite{sandvik1999stochastic}.
\begin{align}
    G_{i,j}(\tau)&=\langle e^{\tau H} S_i^z e^{-\tau H} S_j^z \rangle,\\
    G_k(\tau)&=\frac{1}{L} \sum_{i,j} G_{i,j}(\tau) e^{-i (r_i-r_j)k/L}.
\end{align}
$G_{i,j}(\tau)$ is time-displaced spin-spin correlation of $z$-component between spin $i$ and spin $j$. Target correlation function $G_k(\tau)$ in wave-vector domain is then calculated via Fourier transformation. $r_i$ denotes the location of spin $i$, where the lattice constant is set to be 1. $J$ is used as the energy unit. We set the inverse temperature $\beta=1$ in the Monte Carlo simulation. In this work we'll focus on $k=\pi$ and $G_k(\tau)$ will be represented by $G(\tau)$ for the sake of simplicity. Then the AC task reads $G(\tau)=\int d\omega e^{-\tau \omega} A(\omega)$, where $A(\omega)$ is the target dynamic structure factor. The corresponding sum rule is obtained by setting $\tau=0$, \textit{i.e.}, $\int d\omega A(\omega) = G(0)$.

The same NN structure and hyper-parameters are used as in the previous section. Frequency $\omega$ takes the range $\omega \in [-10,10]$. The time domain and the frequency domain are discretized into 512 and 1024 pieces respectively as before. The spectrum of Heisenberg chain can be regarded as a sum of $\delta$ functions at zero temperature. These $\delta$ functions broaden as temperature increases. We perform quantum Monte Carlo simulation on a 32-site Heisenberg chain, where $\delta$ functions are dense enough on the required energy scale $\Delta \omega \sim 0.02$ so that a smooth spectrum can be obtained. The stochastic series expansion approach with loop-update\cite{sandvik1999stochastic} algorithm is used in simulation. Spin-spin correlation is measured every 100 update steps so that auto-correlation can be ignored. The covariance matrix $\Sigma$ is measured by $\Sigma_{ij}=[\langle G(\tau_i) G(\tau_j) \rangle -\langle G(\tau_i) \rangle \langle G(\tau_j) \rangle]/(N_s-1)$, where $N_s$ is the number of independent samples.

Spin-spin correlation functions are measured using different number of Monte Carlo samples to create datasets of different noise levels. In this section, noise levels are represented by relative statistical errors of $G(0)$, which takes range from $3.8 \times 10^{-3}$ to $3.6 \times 10^{-2}$. Simulated $G(\tau)$ are divided by corresponding $G(0)$ before being fed into neural networks so that the sum rule is restored to $\int d\omega A(\omega)=1$. Then the ``SoftMax'' layer results in the correct sum rule and the scale of extracted spectra will be recovered accordingly by multiplying with $G(0)$. Correlation functions $G(\tau_i)$ at different imaginary time $\tau_i$ are measured independently to ensure zero time-displaced correlation between $G(\tau_i)$. The obtained covariance matrix $\Sigma$ is then a diagonal matrix since $\langle G(\tau_i) G(\tau_j) \rangle -\langle G(\tau_i) \rangle \langle G(\tau_j) \rangle=0$.

\begin{figure}[htbp]
    \centering
    \includegraphics[width=0.9 \linewidth]{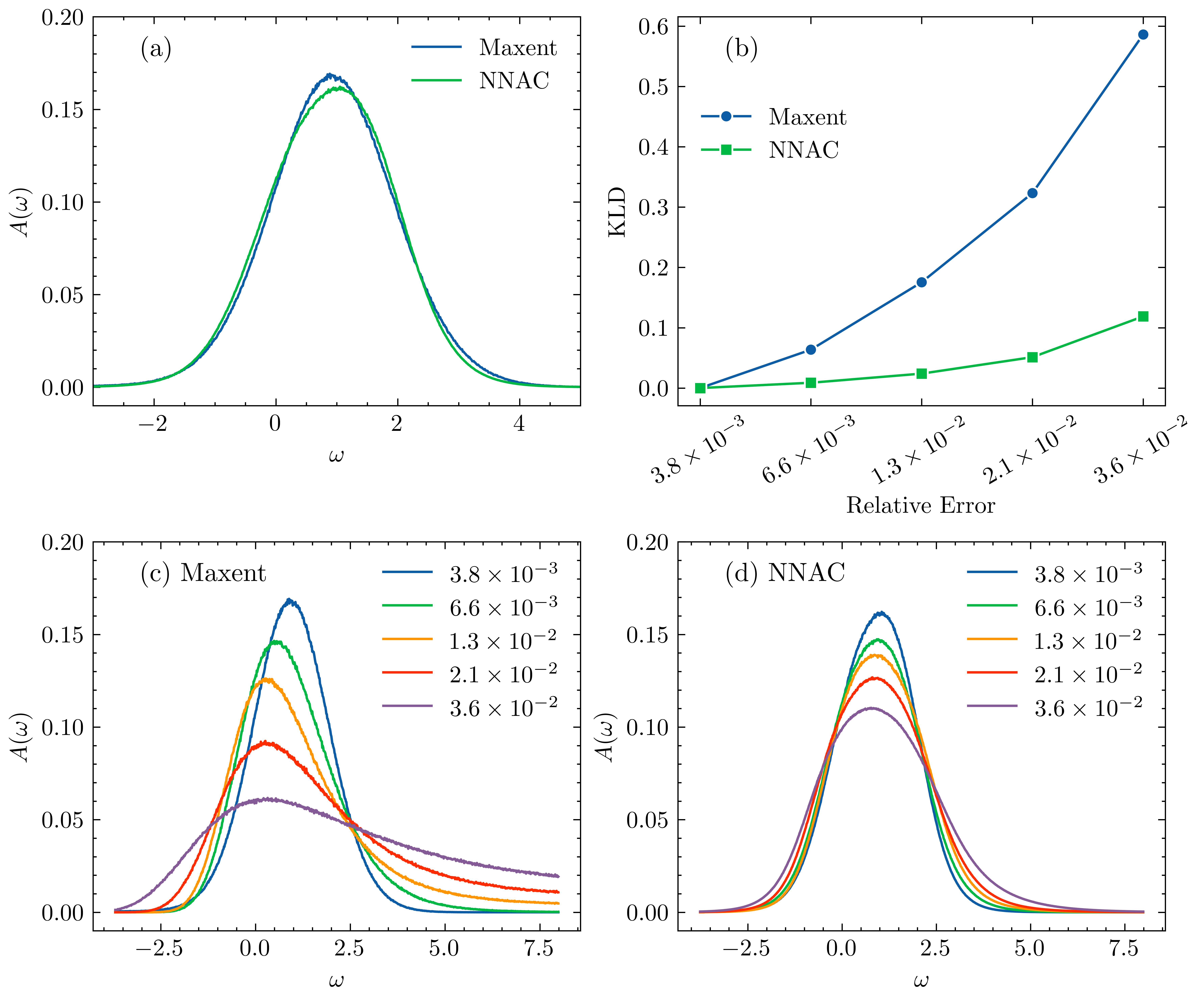}
    \caption{Spectra extracted by different methods. (a) Comparison of spectra generated by Maxent and NNAC from highly accurate $G(\tau_i)$. (b) KLDs of spectra generated by Maxent and NNAC. The most accurate spectra (with the lowest noise level) are taken as ground truths while calculating KLDs. (c) Spectra predicted by Maxent from $G(\tau_i)$ of different noise levels.(d) Spectra predicted by NNAC from $G(\tau_i)$ of different noise levels.}
    \label{Heisenberg}
\end{figure}

Extracted spectra are shown in Figure \ref{Heisenberg}, where Maxent and NNAC are compared. In Figure \ref{Heisenberg} (a), spectra extracted from spin-spin correlation function of relative error $3.8\times 10^{-3}$ by Maxent and NNAC are compared, where two spectra coincide well with each other in this relatively simple single-peak case. These two spectra also agree with those obtained from smaller systems using Lanczos-based methods\cite{okamoto2018accuracy}. Figure \ref{Heisenberg} (b) compares KLDs of the spectra produced by these two methods at different noise levels. Spectra corresponding to the lowest noise level of each method is regarded as ground truths respectively when calculating KLDs. When the noise level increases, the accuracy of the spectra produced by both Maxent and NNAC decreases, but the accuracy of NNAC decays slower than Maxent. Here again, the previous conclusion is confirmed: at low noise levels, Maxnet and NNAC can produce equally accurate results. At high noise levels, however, NNAC performs better than Maxent.

Figures \ref{Heisenberg} (c) and (d) show how spectra extracted by the two methods change when the noise is gradually increased from $3.8 \times 10^{-3}$ to $3.6 \times 10^{-2}$. Spectra get progressively lower and wider in both cases. Spectra generated by Maxent exhibit large peak position shifts, while those generated by NNAC show little shift in peak positions.

\section{Conclusions}
Applications of neural network-based analytic continuation were discussed in this paper.  Numerical experiments are carried on both synthetic datasets and Monte Carlo data. The main conclusion is that a NN can learn from a carefully designed training set and make good predictions on spectra without data leakage, which surpass Maxent in highly noisy cases. To ensure that the neural network acquires adequate knowledge to predict the target spectral functions, the training dataset should comprise a sufficient number of diverse spectral functions. Incorporating information of measured statistical errors leads to better prediction on spectra. $G(\tau_i)$ of the training set should match those of simulated correlation functions in terms of noises at each $\tau_i$ and time-displaced correlation.

While acceptable, the time required for NNAC is relatively long compared to Maxent. Improving the efficiency of model training may be a fruitful area for future investigation. It may be possible to apply the idea of transfer-learning\cite{pan2010survey} here so that we do not need to train a model from scratch for each target spectrum but rather begin with a pre-trained model. A more valuable and ambitious goal is to train a model that is general to any spectrum. The input to this model should probably be the simulated correlation functions and the accompanying covariance matrices, which contain most (if not all) information needed to perform analytic continuation.

\section*{Acknowldgement}
FW acknowledges support from National Natural Science Foundation of China (No. 12274004), and National Natural Science Foundation of China (No. 11888101). Quantum Monte Carlo simulations are performed on TianHe-1A of National Supercomputer Center in Tianjin.

\bibliographystyle{unsrt}
\bibliography{ref}

\end{document}